\documentstyle[psfig,times]{mn}

\def\etal{et~al.}
\def\spose#1{\hbox to 0pt{#1\hss}}
\def\lta{\mathrel{\spose{\lower 3pt\hbox{$\mathchar"218$}}
     \raise 2.0pt\hbox{$\mathchar"13C$}}}
\def\gta{\mathrel{\spose{\lower 3pt\hbox{$\mathchar"218$}}
     \raise 2.0pt\hbox{$\mathchar"13E$}}}

\def\clean{{\sc clean}}

\def\kms{\,km\,s\,$^{-1}$}
\def\Ho50{$H_0 = 50$km\,s$^{-1}$\,Mpc$^{-1}$}

\title[Sub-mm sources towards $z \sim 1$ clusters]
{An excess of sub--millimetre sources towards $\mathbf {z \sim 1}$ clusters} 

\author[P.~N.~Best]{P.~N.~Best,$^1$\thanks{Email: pnb@roe.ac.uk}\\ 
$^1$ Institute for Astronomy, Royal Observatory Edinburgh, Blackford Hill,
Edinburgh EH9 3HJ, UK}

\pagerange{\pageref{firstpage}--\pageref{lastpage}}
\pubyear{2002}
\begin{document}
\label{firstpage}

\maketitle

\begin{abstract}
\noindent Deep sub-millimetre observations using SCUBA are presented of
the central regions of four high redshift clusters which have been
extensively studied optically: CL0023+0423 ($z=0.84$), J0848+4453
($z=1.27$), CL1604+4304 ($z=0.90$) and CL1604+4321 ($z=0.92$).  10
sub--millimetre sources are securely detected towards these four clusters
at 850$\mu$m, with two further tenuous detections; the raw 850$\mu$m
source counts exceed those determined from blank--field surveys by a
factor $\sim 3-4$. In particular, towards CL1604+4304, 6 sources are
detected with $S_{\rm 850 \mu Jy} > 4$\,mJy making this the richest sub-mm
field discovered to date. Corrections for gravitational lensing by these
high redshift clusters reduce these excess sources counts, but are
unlikely to account for more than about half of the excess, with the
remainder presumably directly associated with cluster galaxies. The 450 to
850$\mu$m flux density ratios of the detected sources are systematically
higher (at a significance level $> 98$\%) than those determined for
blank--field selected sources, consistent with them being at the cluster
redshifts.  If subsequent identifications confirm cluster membership,
these results will demonstrate that the optical Butcher--Oemler effect is
also observed at sub-mm wavelengths.
\end{abstract}

\begin{keywords}
submillimetre --- galaxies: clusters: general --- galaxies: starburst
\end{keywords}

\section{Introduction}
\label{intro}

Our understanding of star formation at high redshift has undergone a
revolution in recent years. Lyman--dropout techniques (e.g. Steidel
et~al. 1996)\nocite{ste96} have identified thousands of high redshift
field galaxies with star formation rates of a few $M_{\odot}$\,yr$^{-1}$,
providing estimates of the cosmic evolution of the mean global star
formation rate (e.g. Madau et~al. 1998).\nocite{mad98} However, it has
also become clear that dust plays a very important role in star--forming
galaxies, absorbing a significant proportion of the emitted light.

In order to directly address the role of dust, several sub--mm blank field
surveys have been carried out using the Submillimetre Common--User
Bolometer Array (SCUBA; Holland \etal\ 1999)\nocite{hol99} on the James
Clerk Maxwell Telescope (JCMT), to various depths and sky areas. These
have provided number counts of sub-mm sources down to below 1\,mJy at
850$\mu$m \cite{sma97,hug98b,bar99,bla99b,eal00,sco02,cow02}, and have
resolved the majority of the sub--mm background radiation into
ultraluminous galaxies with SFR $> 100 M_{\odot}$/yr \cite{bla99b}. The
host galaxies of these sources show high optical obscuration, and so are
missed by ultraviolet selection techniques (cf. Adelberger \& Steidel
2000)\nocite{ade00}. SCUBA observations pointed at Lyman--break galaxies
have failed to detect the vast majority to a combined sample noise level
of about 0.5\,mJy, indicating that these galaxies don't contribute
significantly to the sub-mm background (e.g. Chapman \etal\
2000)\nocite{cha00b}. Using a statistical analysis, Peacock \etal\
\shortcite{pea00} demonstrated that Lyman--break galaxies are correlated
with faint background fluctuations (at the $\sim 0.1$mJy level) in a deep
SCUBA image of the Hubble Deep Field, and calculated that on average the
true star formation rates of Lyman--break galaxies are about a factor of 6
higher than those determined from their measured ultraviolet
luminosities. This is consistent with other determinations that the global
star formation rate at high redshifts should be at least a factor of a few
higher than that measured optically (e.g. Hughes et al
1998).\nocite{hug98b}

The ultraviolet and sub--mm selection techniques clearly sample different
populations of star--forming galaxies, and so to obtain a complete census
of star formation at high redshifts, observations of both normal and dusty
galaxies are required.  Whilst these have been carried out for the field,
to date high redshift cluster environments have been neglected at sub--mm
wavelengths.

Rich galaxy clusters contain large numbers of galaxies at the same
distance, and so provide important testbeds for models of galaxy
evolution. The cores of low redshift clusters are dominated by a
population of luminous early-type galaxies; the optical properties of
these evolve only very slowly with redshift, in a manner consistent with
passive evolution of stellar populations which formed at a very early
cosmic epoch (e.g. van Dokkum \etal\ 1998a,b, Schade, Barrientos \&
Lopez-Cruz 1997, Ellis \etal\ 1997, Stanford, Eisenhardt \& Dickinson
1998).  \nocite{sch97,ell97b,dok98b,dok98a,sta98} Low redshift clusters
have been targetted at sub-mm wavelengths, using them as lenses to magnify
the background population; only 2 cluster galaxies have been detected
within 7 clusters at redshifts $0.19 < z < 0.41$ (Edge \etal\ 1999; see
also Chapman \etal\ 2002)\nocite{edg99,cha02}, and both of these were
centre cluster galaxies hosting powerful radio sources, in clusters with
strong cooling flows.

The composition of clusters is a strong function of redshift, however:
Butcher \& Oemler \shortcite{but78} showed that at $z > 0.3$ a population
of bluer galaxies appears in many, but not all, clusters (the
Butcher--Oemler effect). These blue galaxies often exhibit strong
H$\alpha$ or [OII]~3727 emission lines, indicative of on--going star
formation (e.g. Dressler et~al. 1997, 1999).\nocite{dre97,dre99} High
redshift clusters also show an increased number of weak radio sources
(e.g. Dwarakanath \& Owen 1999, Morrison 1999, Smail \etal\ 1999, Best
\etal\ 2002), whose presence is correlated with the Butcher-Oemler blue
galaxy population, though without correspondence of the host galaxies. The
host galaxies of these weak radio sources comprise a mixture of starbursts
and AGN.\nocite{dwa99,mor99,sma99b,bes02a} This increase in activity in
high redshift clusters is not unexpected. Low redshift clusters have the
majority of their gas and galaxies in stable virialised orbits with the
galaxies in the cluster centre having been stripped of gas (e.g. Gunn \&
Gott 1972)\nocite{gun72}, switching off star formation. High redshift
clusters, on the other hand, are much younger, still in their formation
process with relatively high galaxy merger rates (e.g. van Dokkum \etal\
1999)\nocite{dok99a} and a plentiful supply of disturbed gas, and thus
provide ideal laboratories to induce both starbursts and AGN.

To fully understand galaxy and cluster evolution, it is important to
investigate the on--going star formation in these high$-z$ clusters using
sub--mm studies as well as the standard optical selection techniques.
This paper presents the first results of a sub--mm survey of the central
regions of high--redshift clusters, designed to address this issue. The
sample selection, observations and data reduction are described in
Section~\ref{obsred}. In Section~\ref{individs} the four clusters are
individually discussed. The nature of the detected sub-mm sources, their
likelihood of being cluster members, and their global properties are
discussed in Section~\ref{results}. Conclusions are drawn in
Section~\ref{concs}. Values for the cosmological parameters of
$\Omega_{\rm M} = 0.3$, $\Omega_{\Lambda} = 0.7$ and $H_0 =
65$\,km\,s$^{-1}$Mpc$^{-1}$ are adopted throughout.
 
\begin{table*}
\caption{\label{scubaobs} Details of the SCUBA observations. For each
cluster the redshift and pointing centre are given, together with the
observation dates. For each observation date, the total number of 64-pt
jiggle pattern repeats (10 jiggle repeats corresponds to about 30 minutes
observing) and average CSO Tau (at 225 GHz) are given.  }
\begin{tabular}{ccccccc}
Cluster     & z & RA & Dec & Obs Date & N$_{\rm int}$ & $\tau_{\rm CSO}$ \\
            &   &\multicolumn{2}{c}{(J2000)}&(dd-mm-yyyy)&(64-pt jigs)&\\
CL0023+0423 & 0.84 & 00 23 52.3 & +04 22 54 & 25-10-2000 & 130 & 0.09 \\
            &      &            &           & 27-09-2001 &  77 & 0.06 \\
            &      &            &           & 30-09-2001 &  50 & 0.10 \\ 
            &      &            &           & 15-11-2001 &  30 & 0.06 \\
            &      &            &           & 08-12-2001 &  90 & 0.06 \\
J0848+4453  & 1.27 & 08 48 34.6 & +44 53 42 & 26-02-2000 &  90 & 0.07 \\
            &      &            &           & 27-02-2000 &  78 & 0.07 \\
            &      &            &           & 28-02-2000 &  70 & 0.05 \\
CL1604+4304 & 0.90 & 16 04 25.1 & +43 04 53 & 02-03-2000 &  65 & 0.05 \\
            &      &            &           & 03-03-2000 &  90 & 0.04 \\
            &      &            &           & 04-03-2000 &  96 & 0.06 \\
            &      &            &           & 21-03-2000 &  70 & 0.04 \\
            &      &            &           & 22-03-2000 &  70 & 0.03 \\
CL1604+4321 & 0.92 & 16 04 31.5 & +43 21 17 & 25-10-2000 &  20 & 0.09 \\
            &      &            &           & 28-01-2001 &  70 & 0.05 \\
            &      &            &           & 29-01-2001 &  20 & 0.05 \\
            &      &            &           & 04-05-2001 &  40 & 0.08 \\
            &      &            &           & 05-05-2001 &  50 & 0.07 \\
            &      &            &           & 06-05-2001 &  35 & 0.08 \\
            &      &            &           & 08-05-2001 &  30 & 0.08 \\
            &      &            &           & 27-09-2001 &  61 & 0.06 \\
\end{tabular}
\end{table*}

\section{Observations and data reduction}
\label{obsred}

\subsection{Details of the observations}
\label{obsdet}

Cluster targets were selected from the literature based upon the following
criteria: (i) they should have redshifts $z > 0.8$; (ii) deep imaging of
the field should have been carried out, confirming a large overdensity of
red galaxies; (iii) there should exist detailed spectroscopic observations
of the field, with at least 10 galaxies being spectroscopically confirmed
as cluster members. These criteria should ensure that the cluster targets
are indeed genuine high redshift clusters. Four clusters were selected
according to these criteria and observing time constraints: CL0023+0423
($z=0.84$), J0848+4453 ($z=1.27$), CL1604+4304 ($z=0.90$) and CL1604+4321
($z=0.92$). This sample of clusters is not intended to be complete, but
should be representative.

The central regions of the four clusters were observed at 850 and
450$\mu$m simultaneously using SCUBA \cite{hol99} on the JCMT during a
variety of nights between February 2000 and December 2001. Details of the
observations are provided in Table~\ref{scubaobs}. The 850$\mu$m
observations cover an area of sky approximately 2.5 arcminutes in
diameter, and have an angular resolution of about 14 arcsec. At
450\,$\mu$m the observed sky area is slightly smaller and the beamwidth is
about 7.5 arcsec.

Observations were taken using the standard ``jiggle-mode'', whereby the
secondary mirror is moved around 64--point hexagonal pattern, with 1s at
each position, in order to fully sample both the 450 and 850$\mu$m image
planes. The secondary mirror is chopped at 7 Hz between `on--source' and
`off--source' locations, and the telescope is also nodded every 16 seconds
so that the chop position is placed at the opposite side of the field; the
field is therefore observed in an on--off--off--on pattern. A single
64--point jiggle sequence takes 128 seconds (plus overheads) to complete.

For this program the chop throw was fixed to be 45 arcsec in an east--west
direction. This short chop throw results in the `off' beam positions often
also being within the field of view; where this is the case, the sources
will have associated negative counterparts of half intensity 45 arcsec
east and west of the true source. In addition to indications that a short
chop throw improves sky--subtraction, these negative regions provide an
independent measurement of the source, increasing the signal--to--noise
ratios and source reliabilities.  The expected source density is
sufficiently low that these regions are unlikely to coincide with another
source.

A blazar was observed approximately every hour throughout the
observations, to correct the pointing of the telescope. The pointing
measurements before and after each source observation were used to correct
for any pointing shift during the observation. The sky opacities were
monitored by regular skydip observations throughout the nights. Flux
calibration was determined by observing at least one of the flux
calibrators Mars, Uranus and CRL618 each night, using the same 45 arcsec
chop--throw as the targets. It is estimated that the photometry at
850$\mu$m is accurate to about 5--10\%; at 450$\mu$m the photometric
accuracy is worse (predominantly systematics), with variations up to about
25\%.

\subsection{Data reduction and cleaning}
\label{redclean}

The observations were reduced using the SCUBA software, SURF
\cite{jen98a}. The reference measurements were subtracted from the signal
beams and the bolometers were flatfielded using the standard SCUBA
flatfield. Where CSO Tau data was available and reliable (ie. stable, and
consistent with the skydips) throughout the night, the 850 and 450$\mu$m
opacities were calculated for each observation using the CSO Tau relations
provided by Archibald \etal\ \shortcite{arc00b}: $\tau_{\rm 850} = 4.02
(\tau_{\rm CSO} - 0.001)$ and $\tau_{\rm 450} = 26.2 (\tau_{\rm CSO} -
0.014)$. Where this was not possible, opacities were calculated by
interpolating between the values determined from the preceding and
succeeding skydips.  After extinction corrections were applied, noisy
bolometers, noisy integrations and strong spikes in individual bolometers
were rejected; this removed between 5 and 10\% of the data. The sky
variations were removed by subtracting, for each bolometer at each
time-step, the mean signal over all of the other bolometers excluding
those which had significantly higher than average noise.

The calibrated datasets from the different observing nights were weighted
according to their rms noise levels and combined to produce maps of each
cluster. These maps contain negative sidelobes east and west of each
detected source, as discussed in Section~\ref{obsdet}. To remove these and
to improve the fidelity of the images, the maps were cleaned using a
modified version of the \clean\ algorithm \cite{hog74} commonly used in
radio astronomy. In this method, first a normalised beam-map is created
from observations of CRL618 with the same chop--throw. Then, the highest
peak on the source map is found, and the beam-map scaled and subtracted at
this position to remove 10\% of this peak flux (ie. a loop gain of
0.1). This removes some of the positive detection of a source, and
partially fills in its negative sidelobes. The highest peak is allowed to
be either positive or negative: although negative values are unphysical,
this is important otherwise overcleaning of the map can give rise to
spurious features when only positive noise peaks and not negative ones are
cleaned. This cleaning was repeated for a number (typically 50 to 100) of
iterations, until the highest peak found was within 2.5$\sigma$ of the sky
rms noise.

\begin{figure*}
\begin{tabular}{cc}
\psfig{file=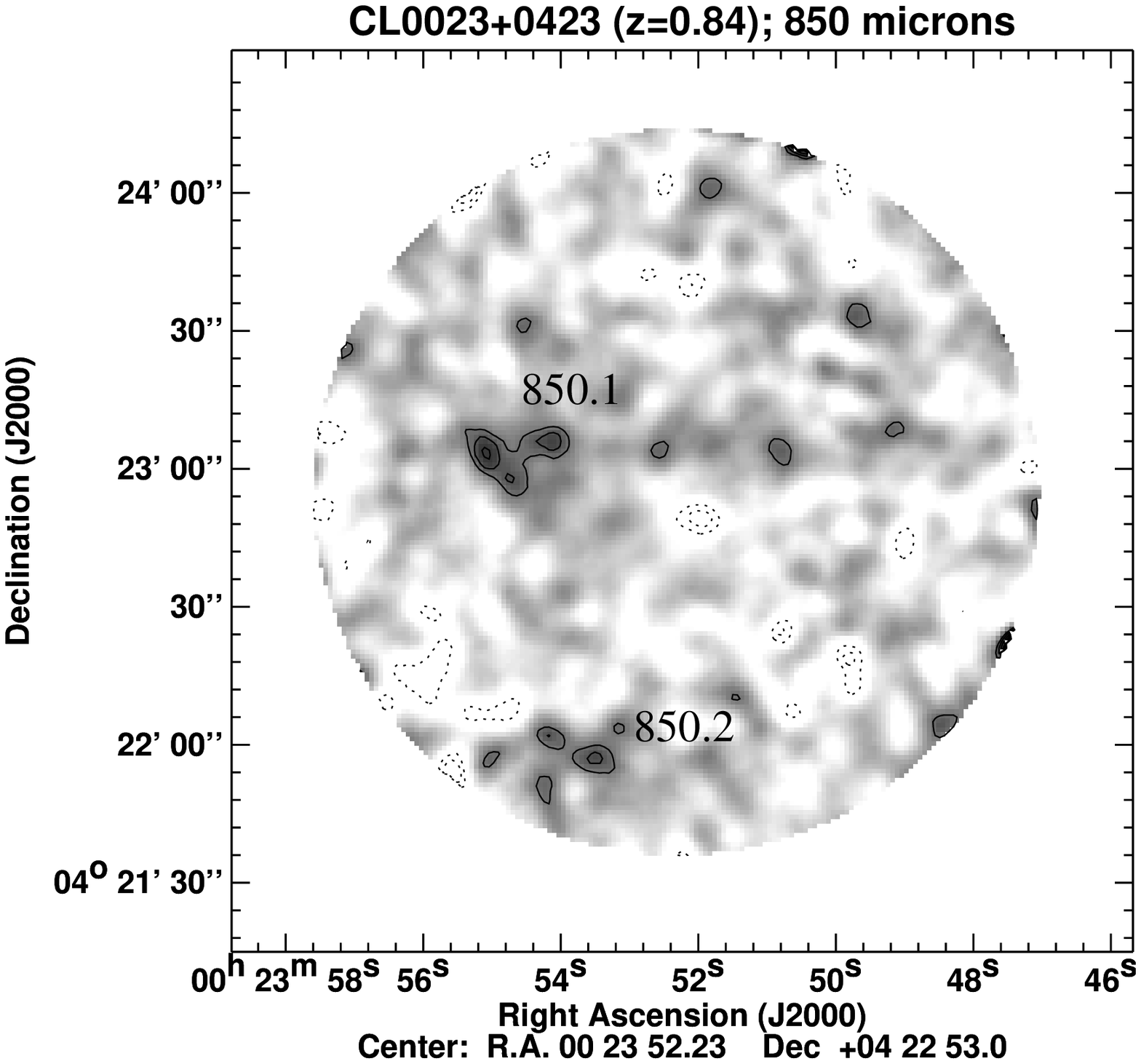,angle=-90,width=8.5cm,clip=}
&						   
\psfig{file=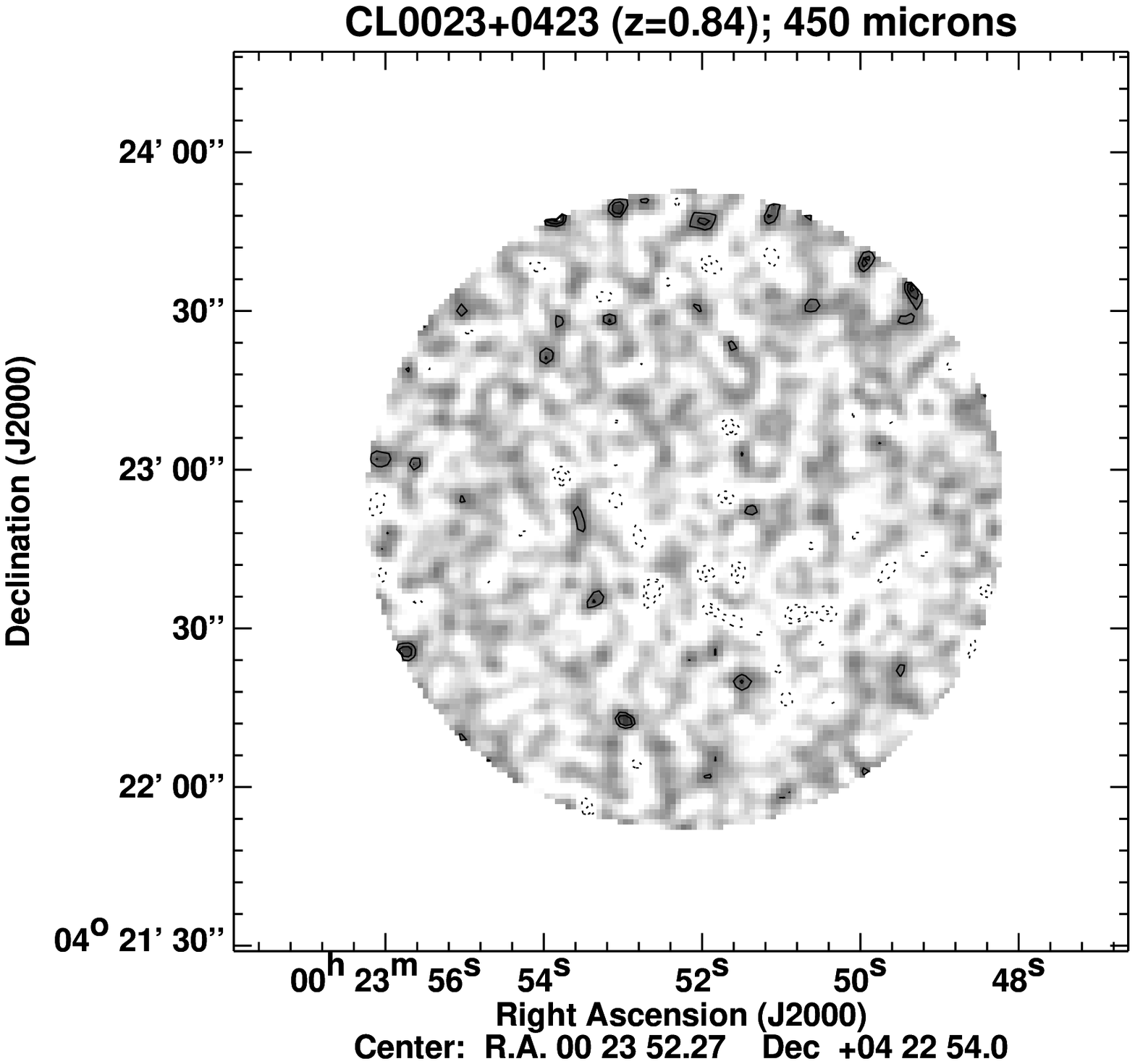,angle=90,width=8.5cm,clip=}
\end{tabular}
\caption{\label{0023figs} SCUBA maps at 850 and 450$\mu$m of the central
regions of the CL0023+0423 cluster at redshift $z=0.84$. Contour levels
are plotted at (-4, -3, 3, 4, 5, 6, 7, 8) times the sky rms noise level
of each map: the 850 and 450$\mu$m maps have rms noise levels of
1.2\,mJy\,beam$^{-1}$ and 9.9\,mJy\,beam$^{-1}$ respectively (note
that the maps have been smoothed).}
\end{figure*}

\begin{figure*}
\begin{tabular}{cc}
\psfig{file=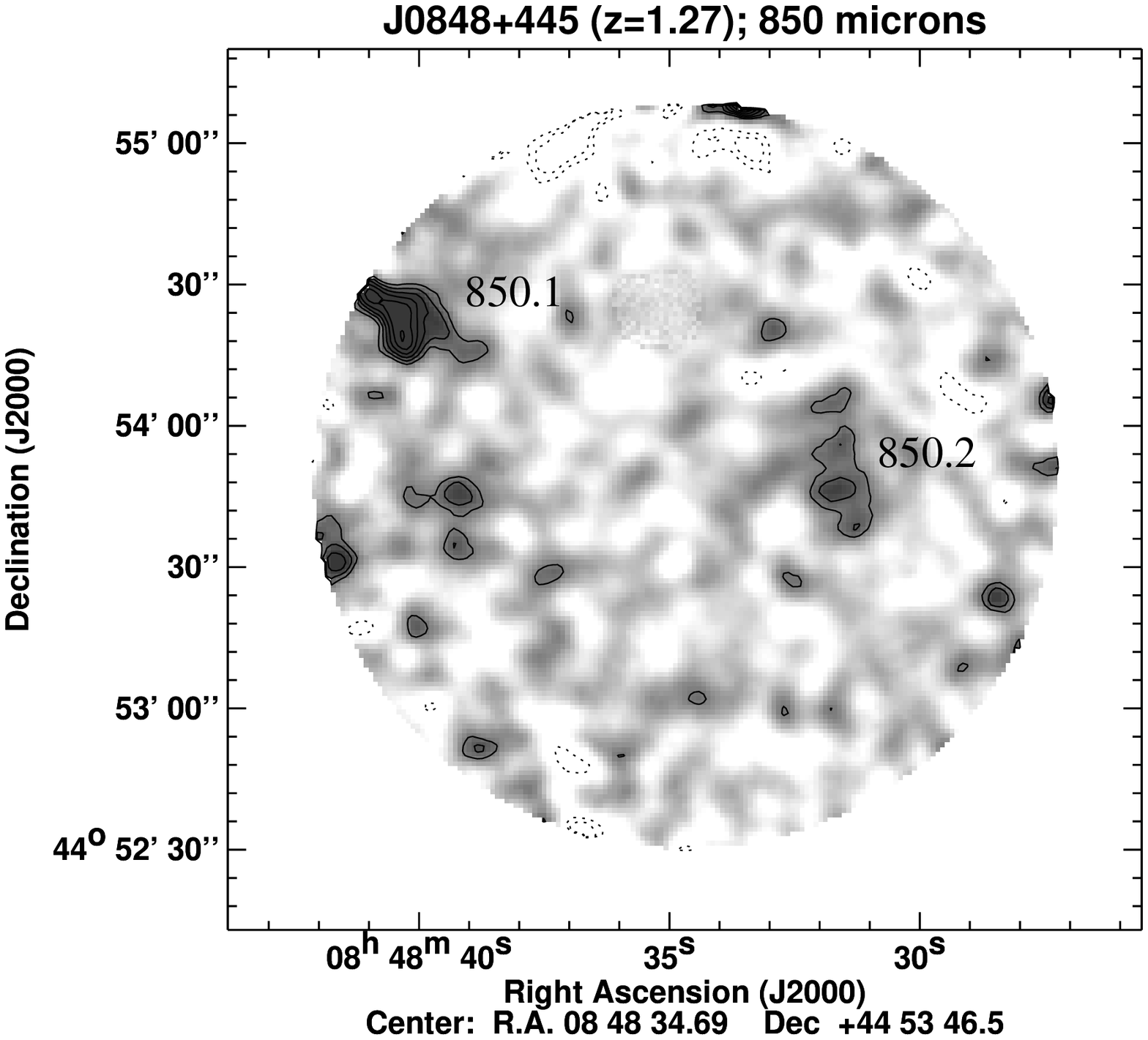,angle=-90,width=8.5cm,clip=}
&		   			 
\psfig{file=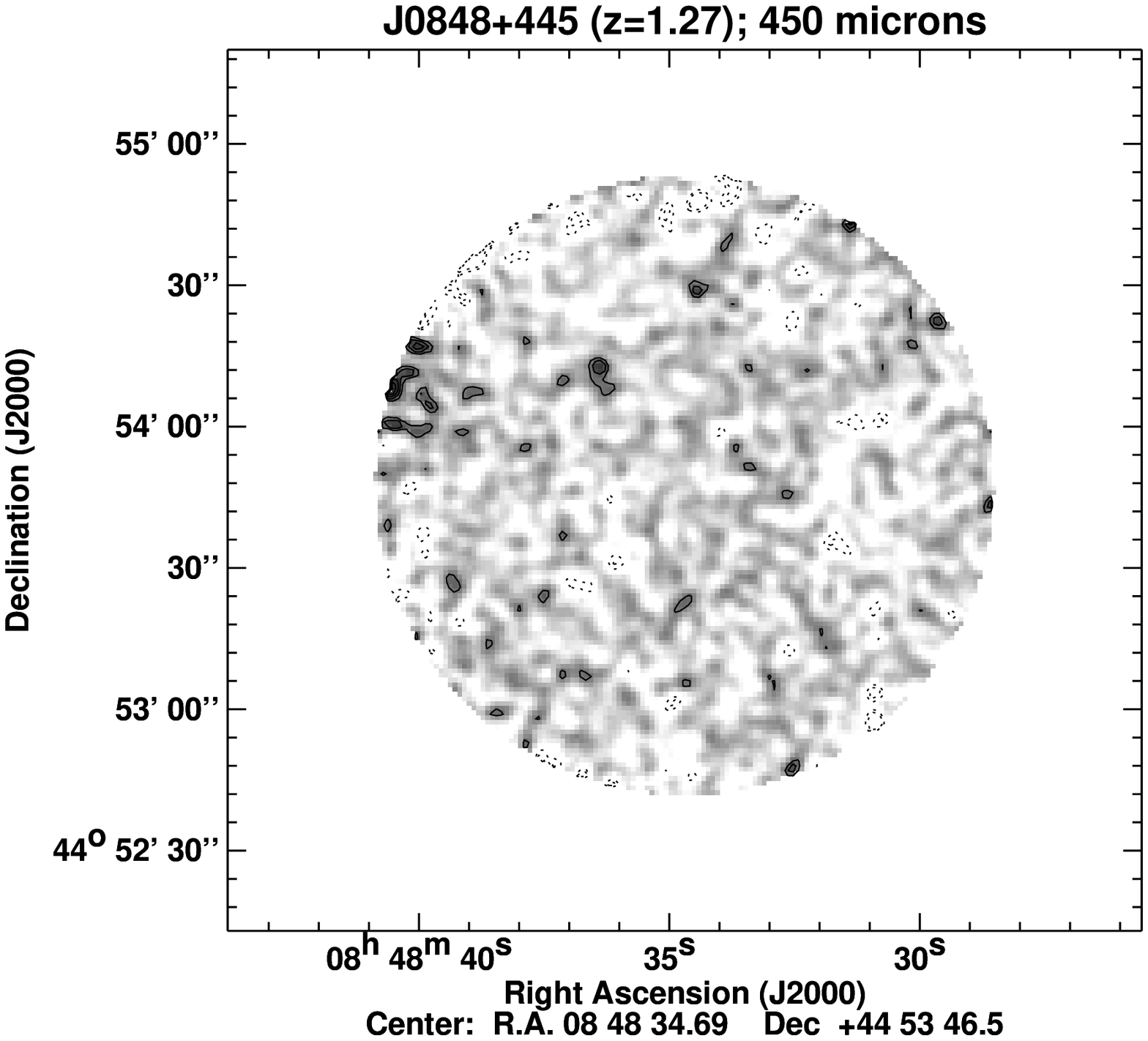,angle=90,width=8.5cm,clip=}
\end{tabular}
\caption{\label{0848figs} SCUBA maps at 850 and 450$\mu$m of the central
regions of the J0848+4453 cluster at redshift $z=1.27$. Contour levels are
plotted at (-4, -3, 3, 4, 5, 6, 7, 8) times the sky rms noise level of
each map: the 850 and 450$\mu$m maps have rms noise levels of
1.6\,mJy\,beam$^{-1}$ and 12.2\,mJy\,beam$^{-1}$ respectively (note
that the maps have been smoothed). A 25 arcsec
diameter region centred on 08 48 35.55, +44 54 24.7 has been masked in
the 850$\mu$m map due to a higher noise level associated with bolometer
flagging.}
\end{figure*}

\begin{figure*}
\begin{tabular}{cc}
\psfig{file=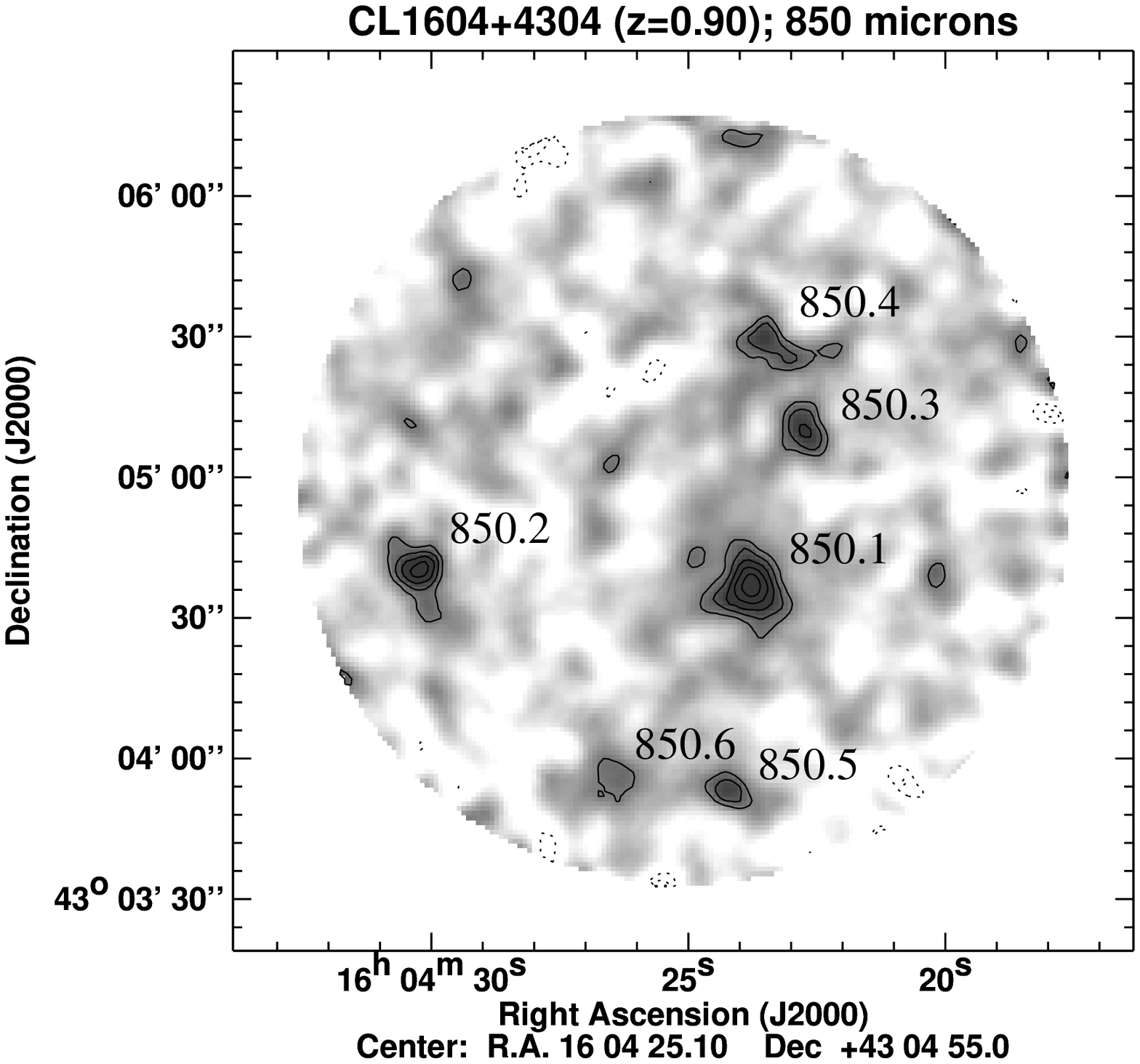,angle=-90,width=8.5cm,clip=}
&						   
\psfig{file=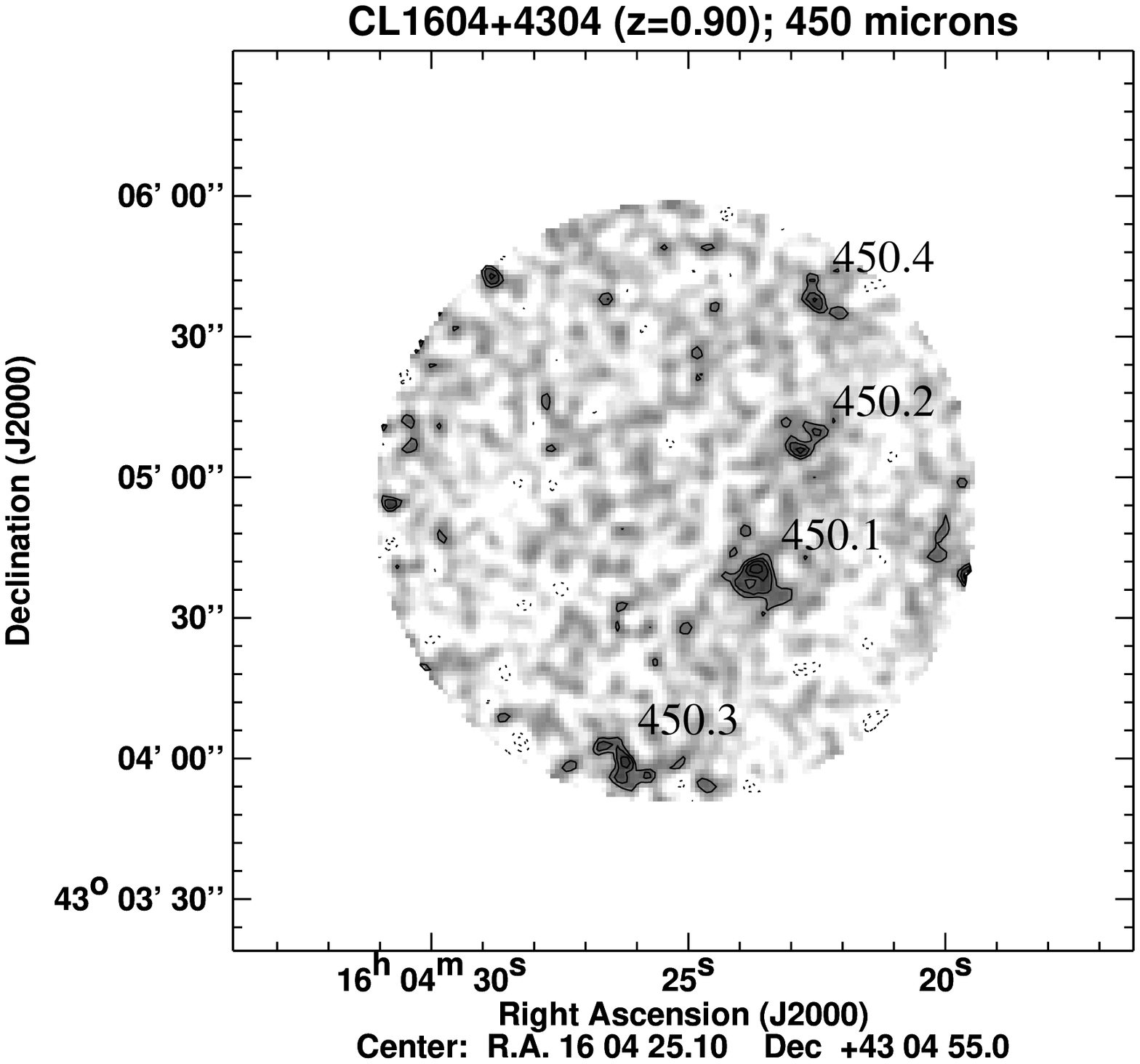,angle=-90,width=8.5cm,clip=}
\end{tabular}
\caption{\label{1604afigs} SCUBA maps at 850 and 450$\mu$m of the central
regions of the CL1604+4304 cluster at redshift $z=0.90$. Contour levels
are plotted at (-4, -3, 3, 4, 5, 6, 7, 8) times the sky rms noise level
of each map: the 850 and 450$\mu$m maps have rms noise levels of
1.1\,mJy\,beam$^{-1}$ and 5.5\,mJy\,beam$^{-1}$ respectively (note
that the maps have been smoothed).}
\end{figure*}

\begin{figure*}
\begin{tabular}{cc}
\psfig{file=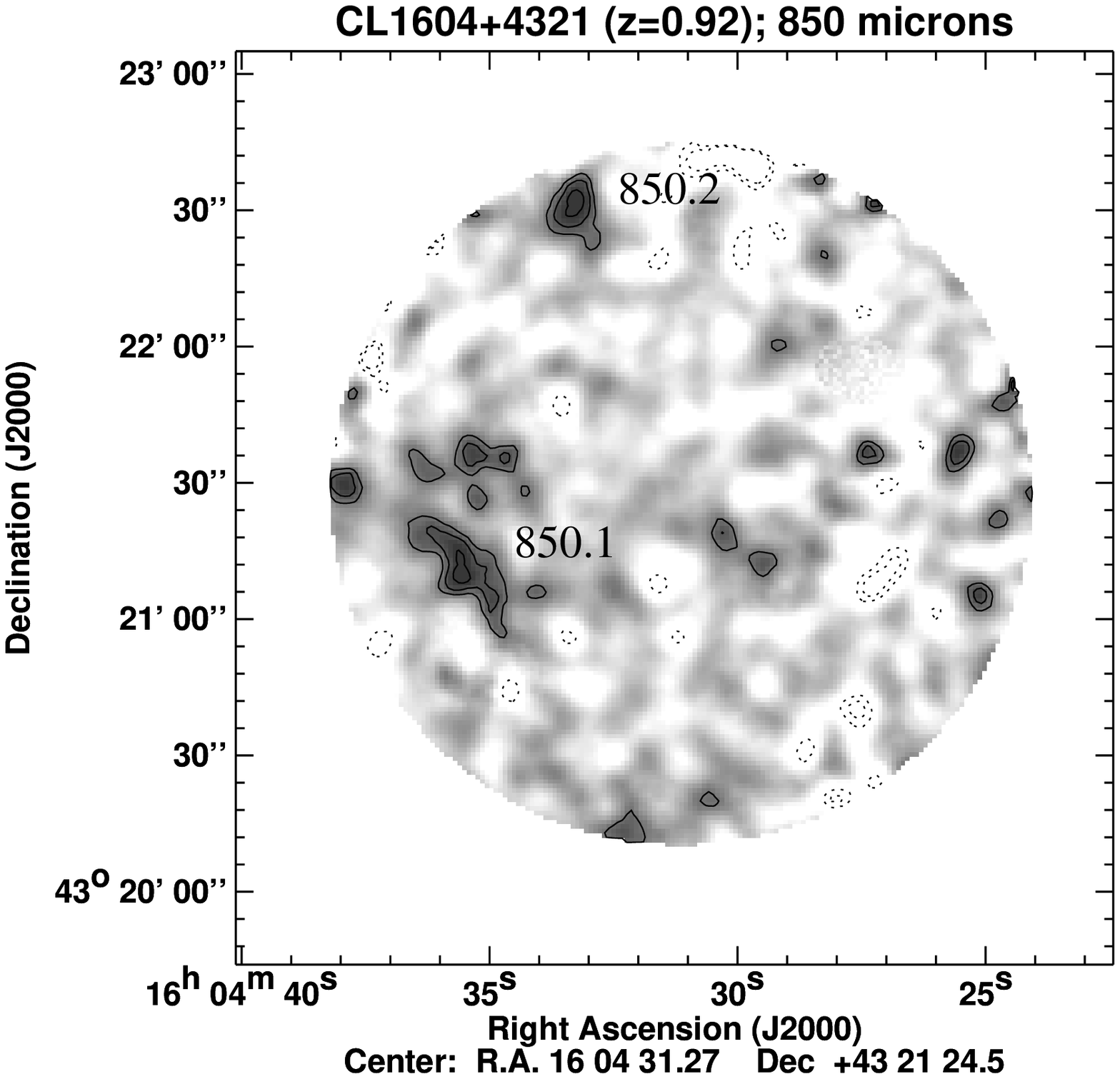,angle=-90,width=8.5cm,clip=}
&
\psfig{file=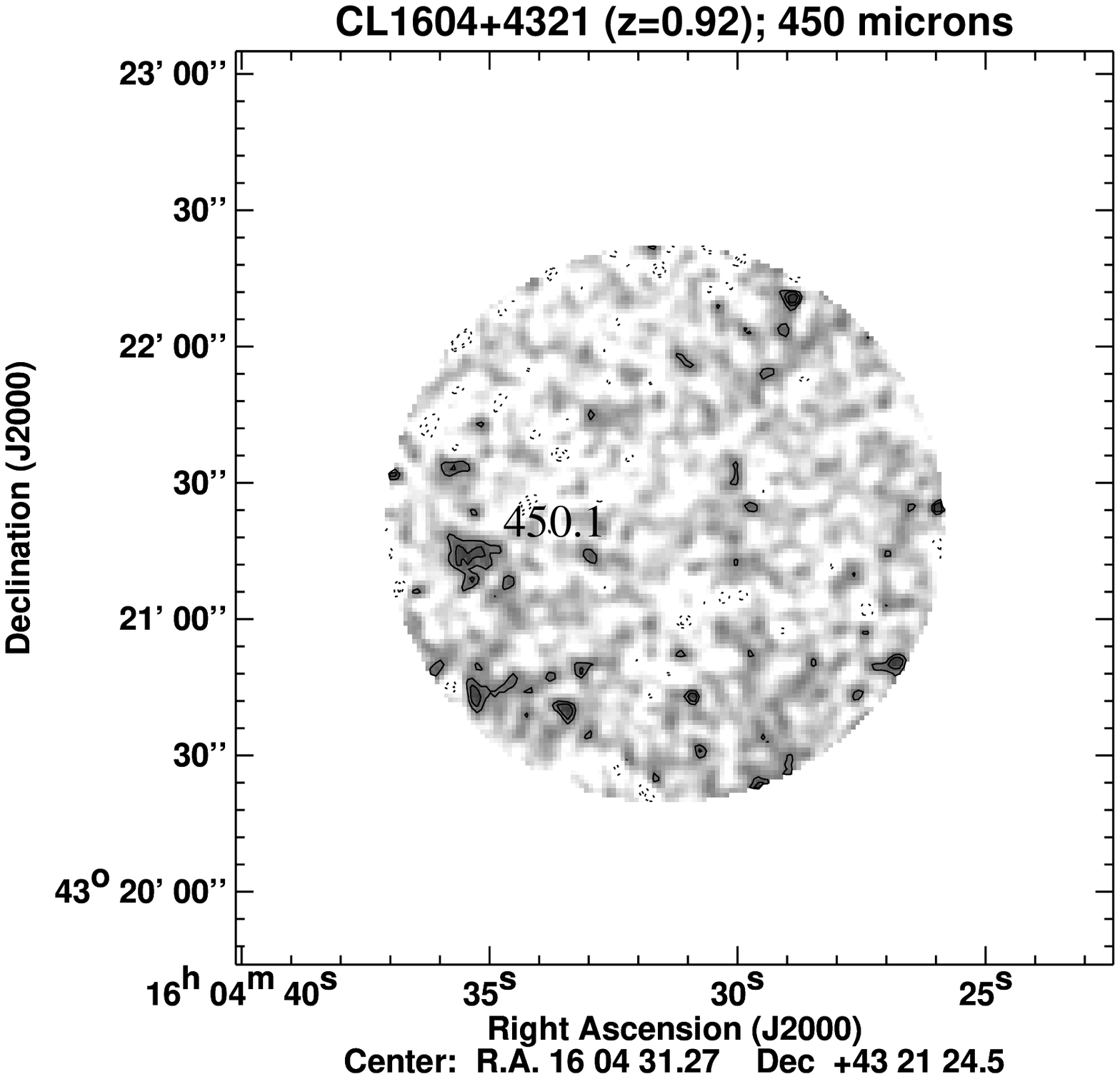,angle=-90,width=8.5cm,clip=}
\end{tabular} 
\caption{\label{1604bfigs} SCUBA maps at 850 and 450$\mu$m of the central
regions of the CL1604+4321 cluster at redshift $z=0.92$. Contour levels
are plotted at (-4, -3, 3, 4, 5, 6, 7, 8) times the sky rms noise level of
each map: the 850 and 450$\mu$m maps have rms noise levels of
1.2\,mJy\,beam$^{-1}$ and 11.1\,mJy\,beam$^{-1}$ respectively (note that
the maps have been smoothed). A 25 arcsec diameter region centred on 16 04
27.63, +43 21 52.9 has been masked in the 850$\mu$m map due to a higher
noise level associated with bolometer flagging.}
\end{figure*}

Each residual map comprises sky noise plus sources fainter than the
minimum cleaned flux density. The flux removed from this map consists of a
set of delta functions around the original source peaks; these are then
convolved with a Gaussian of the same resolution as the JCMT beam and
added back on to the map. The combined maps were then smoothed slightly by
convolving them with a Gaussian, to lower the resolution to 16.5 arcsec at
850$\mu$m and 9.5 arcsec at 450$\mu$m in order to enhance the
detectability of faint sources. The 850$\mu$m and 450$\mu$m maps of the
clusters so produced are displayed in Figures~\ref{0023figs}
to~\ref{1604bfigs}.  Comparison with the initial raw maps confirms that
the cleaning process has reduced the background rms of the maps by
removing the negative sidelobes, but not given rise to any new sources.

\subsection{Source extraction}
\label{sourceext}

Perhaps the most uncertain aspect of the analysis of sub--millimetre data
is the determination of the believability of sources from the final
map. This arises primarily because of the small number of bolometers
across the field (37 for the long wavelength array): noisy (or flagged)
bolometers therefore lead to regions of higher noise in the map and can
give rise to spurious sources. Several steps have been taken to avoid
this. Firstly, all fields have been observed either close to transit or
with some observations either side of transit so that the bolometers
rotate around the field, thus minimising the effects of any bad
bolometers.  Secondly, except for J0848+4453, all fields have been
observed during at least two separate runs, with consequent different
noisy bolometers and different sky coverage. Thirdly, these different
observing periods have been used to test the repeatability of detection of
plausible sources (see below). Fourthly, on two of the 850$\mu$m maps
(J0848+4453 at location 08 48 35.55, +44 54 24.7, and CL1604+4321 at
location 16 04 27.63, +43 21 52.9) the removal of bad bolometer data has
led to regions of significantly shorter exposure and hence higher noise
which may give rise to spurious sources: a 25 arcsec diameter region has
been masked around these two locations.

From the final cleaned maps, both the peak and the integrated flux
densities of likely sources were determined, the latter being measured by
summing the flux density inside an aperture of radius 15 (10) arcsec at
850 (450)\,$\mu$m. The peak flux density (in mJy\,beam$^{-1}$) should be
equal to the integrated flux density of the source (in mJy) for an
unresolved source, and for extended sources it provides a lower limit. In
low signal--to--noise observations, however, the peak flux density can
overestimate the true flux densities, and the integrated flux density
usually provides a more accurate measure. Sources with both peak and
integrated flux densities measured at greater than 3.5$\sigma$, where
$\sigma$ is the rms sky noise of the map, were considered possible
detections. Although within the 8 maps there are negative regions which
reach $-4\sigma$ in peak flux density, in no case does a region reach
$-3.5\sigma$ in both peak and integrated flux, and so there are unlikely
to be any false positive detections at this significance level.

For each field, the observations were then divided into two or three
subsets; in the case of J0848+4453 this was done by separating the first
half of each night from the second half (ie. by bolometer location on the
map), and in the other three cases the separation was made by observing
date.  Independent maps were made for each subset of data. These maps were
then examined to investigate the reliability of the sources detected
above. All of the sources detected above 4$\sigma$ in the overall map were
also present above $\sim 2\sigma$ in each of the data subsets.  

The detected sources are listed in Table~\ref{sources}. The reliability of
each source is also indicated in the table: those sources which are
detected above 4$\sigma$ in integrated flux density are considered secure,
whilst sources between 3.5 and 4$\sigma$ are less so.

\begin{table*}
\caption{\label{sources} The SCUBA sources detected from the maps, at each
of 850 and 450$\mu$m. For each cluster the map rms noise is provided (note
that the maps have been smoothed), together with a listing of the RA, Dec,
offset from the cluster centre, peak flux density, and integrated flux
density of each source. Statistical errors on the flux densities of
individual sources are the rms of the sky background for the peak flux
densities and approximately 1.5 times this value for the integrated flux
densities. In addition there are associated calibration uncertainties for
each map, at a level of 5--10\% at 850$\mu$m and up to 25\% at 450$\mu$m.
The penultimate column of the table indicates for each source whether the
detection is fully secure: sources which are detected above 4$\sigma$ in
integrated flux density are considered secure, whilst sources between 3.5
and 4$\sigma$ are less so. The 3.9$\sigma$ 850$\mu$m source towards
CL1604+4304 is also classified as secure because it is detected at
450$\mu$m. The final column gives the 450/850$\mu$m flux density ratio (or
upper limit) for 850$\mu$m sources with secure detections which lie within
the 450$\mu$m field of view.}
\begin{tabular}{ccccccccccc}
Cluster     & Array & Map rms & Source & RA & Dec & Offset & $S_{\rm peak}$ & 
$S_{\rm int}$ & Secure? & $S_{\rm 450}$/$S_{\rm 850}$ \\
&&[mJy\,/\,bm]& & \multicolumn{2}{c}{(J2000)}& [arcsec] & 
[mJy\,/\,bm]& [mJy]& & ratio \\
CL0023+0423 & 850$\mu$m & 1.2  & HzCl0023\_850.1  & 00 23 54.70 & 
 04 23 02.0 & 37 & 6.6  & 7.4  & Y & $\lta 5.4$  \\ 
            &           &      & HzCl0023\_850.2  & 00 23 53.66 & 
 04 21 55.6 & 62 & 5.6  & 4.7  & ? &\\
            & 450$\mu$m & 9.9  &     ---          &  ---        &
    ---     & ---& ---  & ---  &   &\\
J0848+4453  & 850$\mu$m & 1.6  & HzCl0848\_850.1  & 08 48 40.38 &
 44 54 21.0 & 73 & 13.2 & 13.1 & Y &\\
            &           &      & HzCl0848\_850.2  & 08 48 31.78 &
 44 53 45.0 & 30 & 7.2  & 7.7  & Y & $\lta 6.3$  \\
            & 450$\mu$m & 12.2 &     ---          &  ---        &
    ---     & ---& ---  & ---  &   &\\
CL1604+4304 & 850$\mu$m & 1.1  & HzCl1604a\_850.1 & 16 04 23.93 &
 43 04 35.5 & 22 & 8.5  & 9.4  & Y & $4.6\pm1.4$ \\
            &           &      & HzCl1604a\_850.2 & 16 04 30.38 &
 43 04 37.8 & 60 & 7.3  & 6.4  & Y & $\lta 3.4$  \\
            &           &      & HzCl1604a\_850.3 & 16 04 22.82 &
 43 05 12.1 & 31 & 5.9  & 5.6  & Y & $5.4\pm2.0$ \\
            &           &      & HzCl1604a\_850.4 & 16 04 23.63 &
 43 05 27.8 & 38 & 5.5  & 5.4  & Y & $\lta 4.1$  \\
            &           &      & HzCl1604a\_850.5 & 16 04 24.28 &
 43 03 52.0 & 62 & 5.2  & 4.6  & Y & $\lta 4.8$  \\
            &           &      & HzCl1604a\_850.6 & 16 04 26.38 &
 43 03 55.0 & 60 & 3.9  & 4.3  & Y & $7.0\pm2.8$ \\
            & 450$\mu$m & 5.5  & HzCl1604a\_450.1 & 16 04 23.80 &
 43 04 38.1 & 21 & 36.3 & 43.6 & Y &\\
            &           &      & HzCl1604a\_450.2 & 16 04 22.92 &
 43 05 07.4 & 28 & 29.9 & 30.2 & Y &\\
            &           &      & HzCl1604a\_450.3 & 16 04 26.41 &
 43 03 56.9 & 58 & 30.0 & 30.1 & Y &\\
            &           &      & HzCl1604a\_450.4 & 16 04 22.62 &
 43 05 37.9 & 53 & 27.6 & 19.1 & ? &\\
CL1604+4321 & 850$\mu$m & 1.2  & HzCl1604b\_850.1 & 16 04 35.63 &
 43 21 10.6 & 46 & 6.3  & 6.4  & Y & $7.4\pm2.9$ \\
            &           &      & HzCl1604b\_850.2 & 16 04 33.33 &
 43 22 32.0 & 77 & 6.4  & 4.2  & ? &\\
            & 450$\mu$m & 11.1 & HzCl1604b\_450.1 & 16 04 35.72 &
 43 21 12.8 & 46 & 48.4 & 47.4 & Y &\\  
\end{tabular}
\end{table*}

\section{Discussion of individual sources}
\label{individs}

\subsection*{CL0023+0423}

This cluster was discovered optically and appears in the cluster sample of
Gunn, Hoessel and Oke \shortcite{gun86}. It forms part of the sample of 9
high redshift clusters which Oke, Postman and Lubin \shortcite{oke98} have
studied in detail with deep multi--waveband imaging and spectroscopic
observations. Postman, Lubin and Oke \shortcite{pos98} show that the
optical overdensity is actually the superposition of a cluster at
$z=0.845$, with a velocity dispersion of 415\kms\ (corresponding to a
dynamical mass $M \sim 3 \times 10^{14} M_{\odot}$) and 17
spectroscopically confirmed galaxy members to date, and a group of lower
velocity dispersion ($\sigma \sim 158$\kms) at $z=0.827$, with 7
spectroscopically confirmed galaxies. These two structures are separated
by less than 3000\kms\ in redshift space, and may be interacting: about
57\% of the cluster members show evidence for star formation activity
\cite{pos98}.

The 850$\mu$m SCUBA image of this cluster (Figure~\ref{0023figs}) shows
one source of 7.4\,mJy integrated flux density (HzCl0023\_850.1), which
appears significantly extended; this source appears in all subsets of the
data, although with positional offsets of up to 5 arcsec, presumably
because of the extended nature of the source. A second source
(HzCl0023\_850.2) is detected above 4$\sigma$ in peak flux, but below
4$\sigma$ in integrated flux, and so this is considered a marginal
detection. No believable sources are detected at 450$\mu$m.

\subsection*{J0848+4453}

This cluster, the highest redshift in the current sample, was first
discovered by Stanford \etal\ \shortcite{sta97} as a concentrated
overdensity of very red objects in a near--infrared field survey.  There
is a 4.5$\sigma$ significance X--ray detection ($L_{\rm X} \sim
10^{44}$erg\,s$^{-1}$, 0.5--2\,keV) at the same location.
Follow--up spectroscopy has, to date, confirmed 10 galaxies as cluster
members, with a velocity dispersion of 640\kms\ suggesting a cluster of
about Abell richness class 1 \cite{sta97}. More recently, a second group
of galaxies has been detected only 4 arcminutes away, with redshift 1.26
\cite{ros99}. This group also has associated X-ray emission of similar
luminosity, and spectroscopic redshifts have confirmed cluster membership
for 6 galaxies to date. The J0848+4453 cluster therefore appears to be
part of a massive high redshift structure.

The SCUBA maps of the central regions of this cluster indicate the
presence of two luminous sub-mm sources at 850$\mu$m (see
Figure~\ref{0848figs}; Table~\ref{sources}). A slightly extended source is
seen to the west of centre of the map (HzCl0848\_850.2), and a very
luminous ($>13$\,mJy) source close to the north--eastern edge
(HzCl0848\_850.1).  This latter source is amongst the most luminous
850$\mu$m sources known; although it lies close to the edge of the field,
it occurs in all subsets of the data and so its detection is considered
secure.

\subsection*{CL1604+4304}

Like CL0023+0423, the cluster CL1604+4304 ($z=0.90$) was first discovered
in the optical imaging search of Gunn et~al \shortcite{gun86} and
subsequently formed part of the sample of 9 high--redshift clusters
studied by Oke et~al \shortcite{oke98}. This cluster was studied in
considerable detail by Postman et~al \shortcite{pos98} and Postman, Lubin
and Oke \shortcite{pos01}: spectroscopic redshifts have been obtained for
over a hundred galaxies in the field, with 22 being confirmed cluster
members. The velocity dispersion of these galaxies, $\sigma \sim
1220$\kms, is consistent with an Abell 2 richness cluster although the
cluster is only marginally detected in X-rays with a luminosity of $\sim
10^{44}$\,erg\,s$^{-1}$ (0.1 to 2.4\,keV; Castander \etal\ 1994), and
Smail \etal\ \shortcite{sma94} failed to detect a weak lensing signal
towards it.\nocite{cas94} Approximately half of the confirmed cluster
galaxies are emission line (star forming) objects.

This cluster is particularly interesting as it forms part of one of the
most massive structures known at high redshifts. The equally rich cluster
CL1604+4321 (also in the SCUBA sample: see below) is offset only 17
arcmins on the sky and 4300\,km\,s$^{-1}$ in redshift, and there is a
large overdensity of red galaxies in a `filament' joining the two
clusters, including another smaller condensation of red galaxies
\cite{lub00}. Together these seem to comprise a very massive
high--redshift supercluster.

The 850$\mu$m SCUBA map of the central regions of this cluster is quite
remarkable (Figure~\ref{1604afigs}). This is both the deepest map in the
sample, and the one taken in the best, most stable, sky conditions and
therefore of high image integrity. Five sources are clearly and securely
detected with flux densities in excess of 4.4\,mJy (4$\sigma$; see
Table~\ref{sources}). A sixth source (HzCl1604a\_850.6) is detected
slightly below 4$\sigma$ significance, but appears as a luminous source in
the 450$\mu$m observations (HzCl1604a\_450.3), confirming its reality. A
further two of the 850$\mu$m sources are also detected at 450$\mu$m,
providing still further confirmation of their reality. A fourth 450$\mu$m
source is marginally detected at between 3.5 and 4$\sigma$, but the
disparity between its peak and integrated flux densities and its
non--detection at 850$\mu$m cast doubt on its reality. The rms of the
450$\mu$Jy map is significantly lower than those of the other clusters in
the sample due to the low opacities during these observations.

The detection of 6 sources brighter than 4\,mJy at 850$\mu$m makes this
the richest sub-mm field observed to date. The source density on this map
is a factor of 5 higher than expected from blank field counts (see
Section~\ref{counts}).
 
\subsection*{CL1604+4321}

As discussed above, CL1604+4321 at redshift 0.92 forms part of a potential
high redshift supercluster together with CL1604+4304. Postman \etal\
\shortcite{pos01} studied CL1604+4321 in detail, and spectroscopically
confirmed about 40 galaxies as being associated with the cluster, with a
velocity dispersion of 935\kms.

One source is clearly detected above the 4$\sigma$ level at both 850 and
450$\mu$m in the SCUBA maps (HzCl1604b\_850.1 and 450.1;
Figure~\ref{1604bfigs}; Table~\ref{sources}).  A second source, towards
the north--east edge of the 850$\mu$m map, is detected with an integrated
flux density level between 3.5 and 4$\sigma$, and so is considered to be
an insecure detection.

\section{Results}
\label{results}

Ten sub-mm sources have been securely detected towards these four high
redshift clusters at 850$\mu$m, with a further two possible detections.
The important issue is now to determine whether these sources are indeed
associated with the clusters or whether they are merely projected along
the same line of sight. Before all of the sources are identified and their
redshifts measured, this question cannot be unambiguously answered, but in
this section some indications are given based upon the source counts and
upon the 450\,/\,850$\mu$m flux density ratio which serves as a rough
redshift indicator.

\subsection{Integrated Source counts}
\label{counts}

The integrated source counts at 850$\mu$m and 450$\mu$m have been
determined as a function of limiting flux density using the integrated
flux densities of each source. Scott \etal\ \shortcite{sco02} found, using
simulated maps, that the flux densities of sources in the 8\,mJy survey
are typically overestimated due the effects of both noise and source
confusion; here no corrections were made for this effect, because our sky
area covered is too small and contains too many sources to accurately
calculate any such corrections. At the significance levels ($> 4\sigma$)
of the sources considered, the flux boosting effects due to noise should
be minimal, and in any case any such flux boosting is compensated by the
fact that the 15 arcsec radius aperture will not contain the total source
flux density --- in theory only $\sim 95$\% for a point source. Confusion
effects, however, could be important if faint sources below the flux limit
combine to give an apparent source above that limit. One or two of the
more extended sources could conceivably be the result of source confusion,
although the simulations of Hughes \& Gazta{\~n}aga \shortcite{hug00b}
suggest that this effect is of little importance at flux densities above
$\sim 2.5$\,mJy, and so from here on it is discounted.

Only the securely detected sources were considered in this analysis. At
850$\mu$m the sky area observed in each map was taken to be 4.5 square
arcminutes down to a flux density level of 4 times the rms noise quoted in
Table~\ref{sources}, plus a further 1 square arcminute around the edges of
the map to a flux density level 50\% higher. At 450$\mu$m the sky areas
were similarly 3 square arcminutes at the highest sensitivity and 1 square
arcminute at the lower sensitivity. Between each of the 8 limiting flux
densities (that is, the limiting values for the central and edge regions
of each map), the sky area sampled brighter than this limit was determine
and used to calculate the number density of sources within that flux
density range. From these were calculated the number density of sources
above three flux density levels at 850$\mu$m and a single level at
450$\mu$m. These are provided in Table~\ref{srccnts}. It is important to
note that at this stage no account is taken of any possible gravitational
lensing by the clusters, and so the raw source counts likely overestimate
the true counts; this is discussed in detail in Section~\ref{lensing}.

\begin{figure*}
\centerline{
\psfig{file=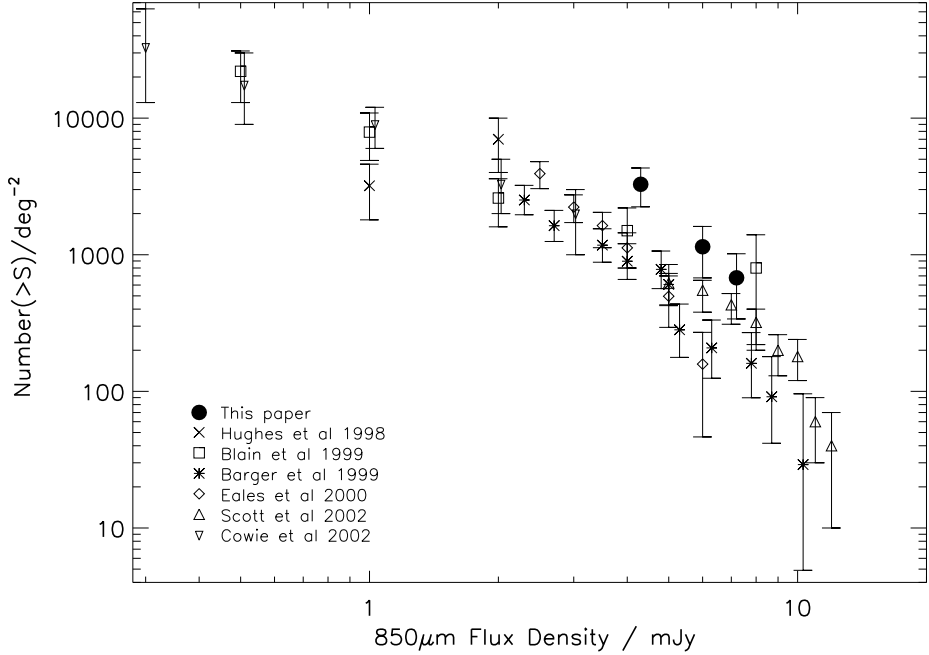,angle=90,width=13cm,clip=}
}
\caption{\label{850cnts} The integrated source counts at 850$\mu$m as
observed in the fields of the high redshift clusters (filled circles)
compared with those determined from various SCUBA field surveys (open
symbols), taken from the references indicated in the lower left corner of
the plot. A clear excess of sub-mm galaxies is seen in the direction of
the high redshift clusters.}
\end{figure*}

The integrated source counts determined at 850$\mu$m are compared with
those determined from blank field surveys on Figure~\ref{850cnts}. The
comparison data are taken from the Hubble Deep Field survey \cite{hug98b},
the various SCUBA lensing surveys \cite{sma97,bla99b,cow02}, the Hawaii
Survey Fields \cite{bar99}, the Canada--UK Deep Submillimetre Survey
\cite{eal00} and the 8\,mJy Survey \cite{sco02}.  Figure~\ref{850cnts}
shows that the 850$\mu$m counts towards the high redshift clusters are
clearly in excess, by a factor $\sim$3--4, of those in the blank field
surveys. Note that a similar excess of 850$\mu$m counts has also been
found in the field surrounding the high redshift ($z=3.8$) radio galaxy
4C14.17, which is thought to reflect a high--redshift proto--cluster
\cite{ivi00b}. It should be noted that the sky areas sampled are
relatively small, and SCUBA sources are known to be highly clustered
\cite{alm02}, which could give rise to higher than average counts being
measured in some locations. Although this possibility cannot be
definitively excluded until identifications and redshifts are available
for the sources, for such a strong overdensity to occur by chance just at
the location of these clusters would be an odd coincidence, and so it is
far more likely that the excess is real (or due to lensing; see below).

At 450$\mu$m determining the source counts is considerably more uncertain
due to the fewer sources detected and the larger uncertainties in
calibration. Smail \etal\ \shortcite{sma02} derive 450$\mu$m source counts
to be $2100 \pm 1200$\,deg$^{-2}$ sources brighter than 10\,mJy and $500
\pm 500$\,deg$^{-2}$ sources brighter than 25\,mJy. The $3800 \pm 1900$
counts deg$^{-2}$ determined here to the same 25\,mJy flux density limit
is almost an order of magnitude in excess of their counts, although the
errors are clearly large.

\begin{table}
\caption{\label{srccnts} The integrated source counts as a function of
limiting flux density at 850 and 450$\mu$m as observed in the fields of
the high redshift clusters.}
\begin{tabular}{cccc}
Wavelength  & Flux density &   N($>$S)   & $\Delta$N  \\
\ [$\mu$m]\ &  [mJy]       &[deg$^{-2}$] &[deg$^{-2}$] \\  
 850        &   7.2        &   680       &    340      \\ 
            &   6.0        &   1150      &    470      \\
            &   4.3        &   3300      &   1000      \\
\\
 450        &   25.0       &   3800      &   1900      \\
\end{tabular}
\end{table}
    
\subsection{Gravitational lensing?}
\label{lensing}

Because these sources are observed in directions towards concentrations of
mass, the possibility of gravitational lensing by the high redshift
clusters must be considered. Gravitational lensing has two competing
effects upon the observed distribution of source counts: the apparent
luminosity of each lensed source is increased, boosting the apparent
source counts, but the sky area observed is decreased in the lensed
regions. 

Prior to any lensing the total source counts observed will be $N(S_{\nu})
= \int_0^{z_{max}} n(S_{\nu},z) {\rm d}z$, where $n(S_{\nu},z)$ describes
the source counts as a function of redshift. After lensing by a factor
$\mu(\theta,z)$, where $\theta$ is the angular offset from the lens
centre, the observed source counts will be

\begin{displaymath}
N'(S_{\nu},\theta) = \int_0^{z_{max}} \frac{1}{\mu(\theta)}
~n\left (\frac{S_{\nu}}{\mu(\theta)},z \right ) {\rm d}z. 
\end{displaymath}

If the source counts can be described by a power law, $n(S_{\nu},z)
\propto S_{\nu}^{-\alpha}$, then $n(\frac{S_{\nu}}{\mu},z) = \mu^{\alpha}
n(S_{\nu},z)$ and therefore $N'(S_{\nu},\theta) = \mu(\theta)^{\alpha-1}
N(S_{\nu})$.  When the slope ($\alpha$) of the power-law exceeds unity
then the effects of gravitational lensing can boost the observed source
counts.  Figure~\ref{850cnts} demonstrates that the sub--mm source counts
indeed have a very steep slope: the slope of the integrated source counts,
$N(>S_{\nu})$, is about 1.8, and so $\alpha \sim 2.8$. Thus a
magnification of a factor 2 provides a boost in the source counts of
approximately a factor 3.5. It is this dramatic effect which has led to
low redshift rich clusters being used as lenses to study the faint sub-mm
source population (e.g. Smail et~al 1997)\nocite{sma97}.

Modelling the lensing object as a singular isothermal sphere (SIS), that
is, one with a central density profile of $\rho(r) \propto r^{-2}$, the
factor by which the sources are magnified at an angular offset $\theta$ is
related to the Einstein radius, $\theta_{\rm E}$, by:

\begin{displaymath}
\mu(\theta) = \left |1 - \frac{\theta_{\rm E}}{\theta} \right |^{-1}
\end{displaymath}

\noindent The Einstein radius can be expressed in terms of the properties
of the source and lens as (e.g. Blandford and Narayan 1992)\nocite{bla92b}: 

\begin{displaymath}
\theta_{\rm E} = \frac{4 \pi \sigma^2}{c^2} \frac{D_{\rm LS}}{D_{\rm OS}}
\end{displaymath}
 
\noindent where $\sigma$ is the velocity dispersion of the lens, $D_{\rm
LS}$ is the angular distance between the source and lens, and $D_{\rm OS}$
is the angular distance between the source and the observer. The lensing
magnification is maximal for sources observed at the Einstein radius
(theoretically infinite for a point source, and in practice giving rise to
strongly--lensed arcs or multiple images) but falls off rapidly beyond
this: at $\theta = 2\theta_{\rm E}$ the magnification factor $\mu(\theta)
\sim 2$, and by 4$\theta_{\rm E}$ it is only a factor of 1.33. For flatter
central density profiles, for example the NFW profile (Navarro, Frenk \&
White 1997) \nocite{nav97} with $\rho(r) \propto r^{-1}$, the
magnification factors beyond $\theta_{\rm E}$ are lower than for the SIS.

The velocity dispersions of the four clusters in this sample are very
uncertain because of the small numbers of redshifts currently available.
The estimated values range from 415 to 1220\kms\ for the four clusters
\cite{ros99,pos01}. Adopting a velocity dispersion of 1000\kms, a lens 
redshift $z_{\rm L}=1$ and a source redshift $z_{\rm S} = 3$, the
Einstein radius corresponds to $\theta_{\rm E} \sim 13$\,arcsec.  This
value is significantly smaller than for low redshift rich clusters (up to
30 or 40 arcsec), firstly because the high redshift clusters are of
significantly lower mass (velocity dispersion) and secondly because the
optimal lensing configuration has equal source--lens and lens--observer
distances which favours $0.2 \lta z_{\rm L} \lta 0.4$ for $2 \lta z_{\rm
S} \lta 4$. Therefore the effects of magnification by these high redshift
clusters are of lesser importance.

A proper determination of the unlensed source counts cannot be carried out
without an accurate mass model for these clusters, but a rough judgement
of the importance of lensing magnification can be made.  Integrating the
expression for the magnified source counts across the field using an
Einstein radius of 13 arcsec ($\sigma \sim 1000$\kms) predicts a total
boosting of the 850$\mu$m source counts of a factor between 1.5 and 2 for
an SIS profile. This is consistent with the boosting in number counts that
might be expected by examining the distribution of detected sources within
the fields: the offsets of each individual source from the cluster centre
are given in Table~\ref{sources}. HzCl1604a\_850.1 has a angular offset
from the cluster centre of only $\sim 20$ arcsec indicating that it is
likely to be significantly magnified; the relatively low 450 to 850$\mu$m
flux density ratio of this source (see Section~\ref{redinds}) further
supports the suggestion that this is a background source. Four further
objects (HzCl0023\_850.1, HzCl0848\_850.2, HzCl1604a\_850.3 and
HzCl1604a\_850.4) have angular offsets of between 30 and 40 arcsec, with
consequent flux boosting of $\sim 50$\% if these are lensed background
objects, but all of the remaining 7 sources are more than 40 arcsec away
from their cluster centres, with expected flux boostings of $\lta
25$\%. The radial distance of all but one sources is at least double the
Einstein radius, making it extremely unlikely that any multiple image
strong lensing affects the number counts in these clusters.

It should be stressed that, unless the current velocity dispersion
measurements greatly underestimate the mass of these clusters, these
calculations represent the upper limit of the boosting that gravitational
lensing by the cluster could feasibly provide. The $\sigma \sim 1000$\kms\
adopted is very much at the maximum for clusters at these redshifts, and
the SIS profile represents the `worst--case' scenario. Further, for
CL1604+4304 with the most detected sub-mm sources and the highest
estimated velocity dispersion, not even a weak--lensing signal was seen
for optical galaxies in the analysis of Smail \etal\ \shortcite{sma94}.
Therefore although cluster gravitational lensing is likely to have a
significant effect upon the source counts, this is not nearly to the
extent of the factor $\sim 3.5$ higher source counts that is observed.

An alternative factor that could conceivable boost the source counts is
gravitational lensing by individual galaxies. It has been suggested that
the brightest SCUBA source in the Hubble Deep Field is gravitationally
lensed by a $z \sim 1$ elliptical \cite{dow99}, and within a $z \sim 1$
cluster there is a higher surface density of elliptical galaxies which
could act as lenses. The Einstein radius of an individual elliptical
galaxy at these redshifts will be only $\theta_{\rm E} \sim 1$--2\,arcsec;
an examination of optical data available for CL1604+4304 \cite{sma94}
suggests that the SCUBA source lies this close to a bright elliptical in
only one of the six cases, that being HzCl1604a\_850.1 which was already
suggested to suffer strong cluster lensing. Once again, therefore, this
effect is unlikely to significantly affect the number count statistics,
but could be relevant in isolated individual cases. This is an important
effect to bear in mind during follow-up studies as it can easily lead to
misidentification of the host galaxy (cf Downes \etal 1999)\nocite{dow99}.

All in all, the conclusion must be (particularly when coupled with the
redshift indicators below) that the higher source counts are in part due
to lensing effects, both from the cluster and individual galaxies, but
that a significant population of cluster sources is also detected.

\subsection{Redshift indicators}
\label{redinds}

For a dust temperature of 30 to 50K, the grey--body dust spectral energy
distribution (SED) peaks at $\sim 100\mu$m. Thus, whilst the 850$\mu$m
data point lies on the Rayleigh--Jeans tail, providing a strong positive
K--correction which results in the 850$\mu$m flux density of a given
source appearing the same at all redshifts $1 \lta z \lta 5-10$
(e.g. Blain \& Longair 1993)\nocite{bla93}, the 450$\mu$m data point moves
steadily through the peak of the SED with increasing redshift. The 450 to
850$\mu$m flux density ratio therefore decreases monotonically with
redshift and so can be used as a broad redshift indicator. 

A note of caution needs to be added when applying this relation in
practice, since 450$\mu$m calibration is notoriously difficult except in
the best atmospheric conditions (such as those for the observations of
CL1604+4304), and calibration errors would lead to systematic offsets, and
hence uncertain results. The radio to sub-mm spectral index (e.g. Carilli
\& Yun 2000)\nocite{car00} is a better tool to use as a redshift
estimator, and in addition the accurate positions determined from the
radio data allow optical identifications to be made. Radio observations of
these fields are proceeding and will provide more definitive results in
the future, but with the current data the 450 to 850$\mu$m flux density
ratio can provide some good first indications.

\begin{figure}
\begin{tabular}{c}
\psfig{file=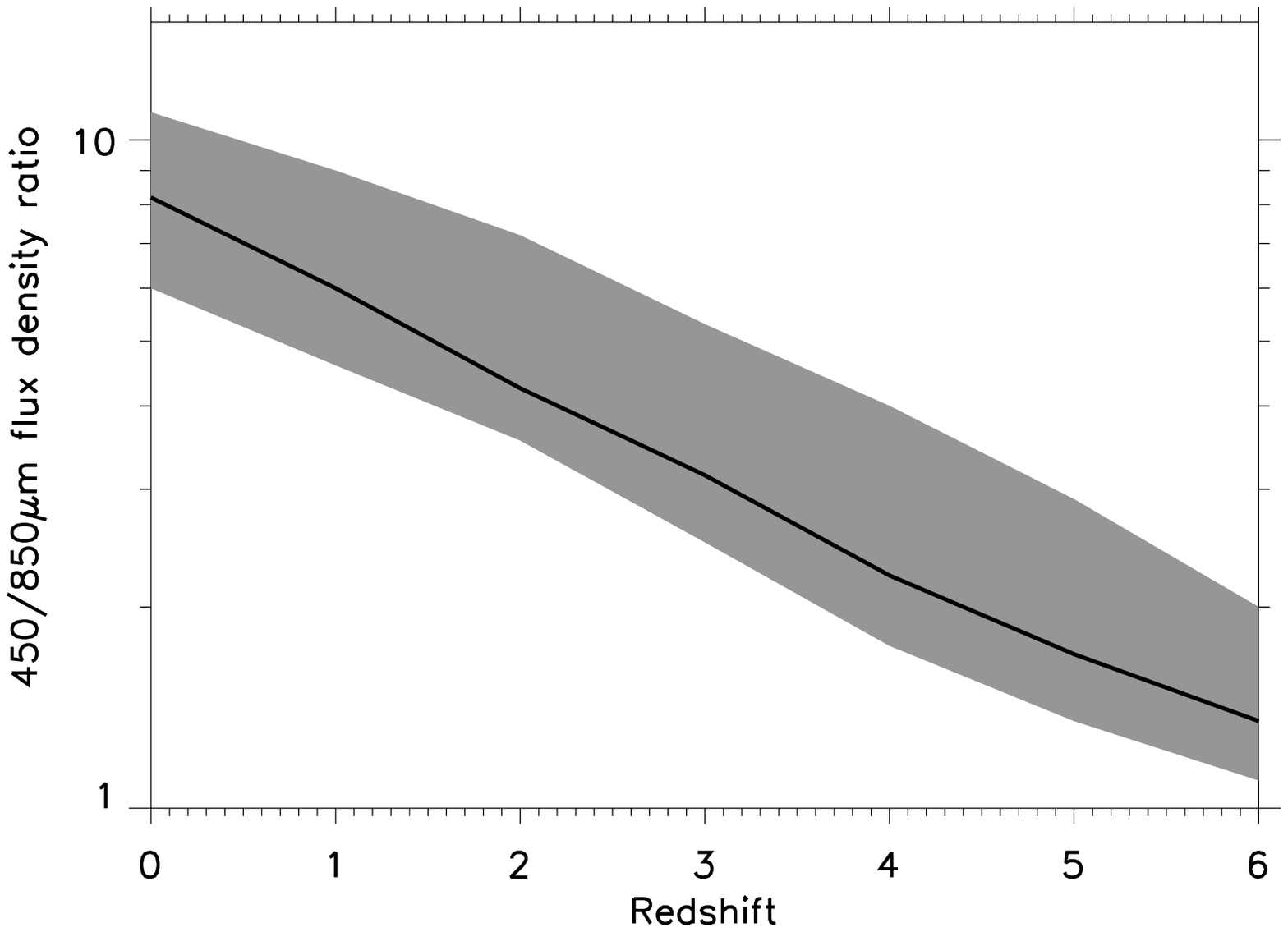,width=8.3cm,clip=}\\
\psfig{file=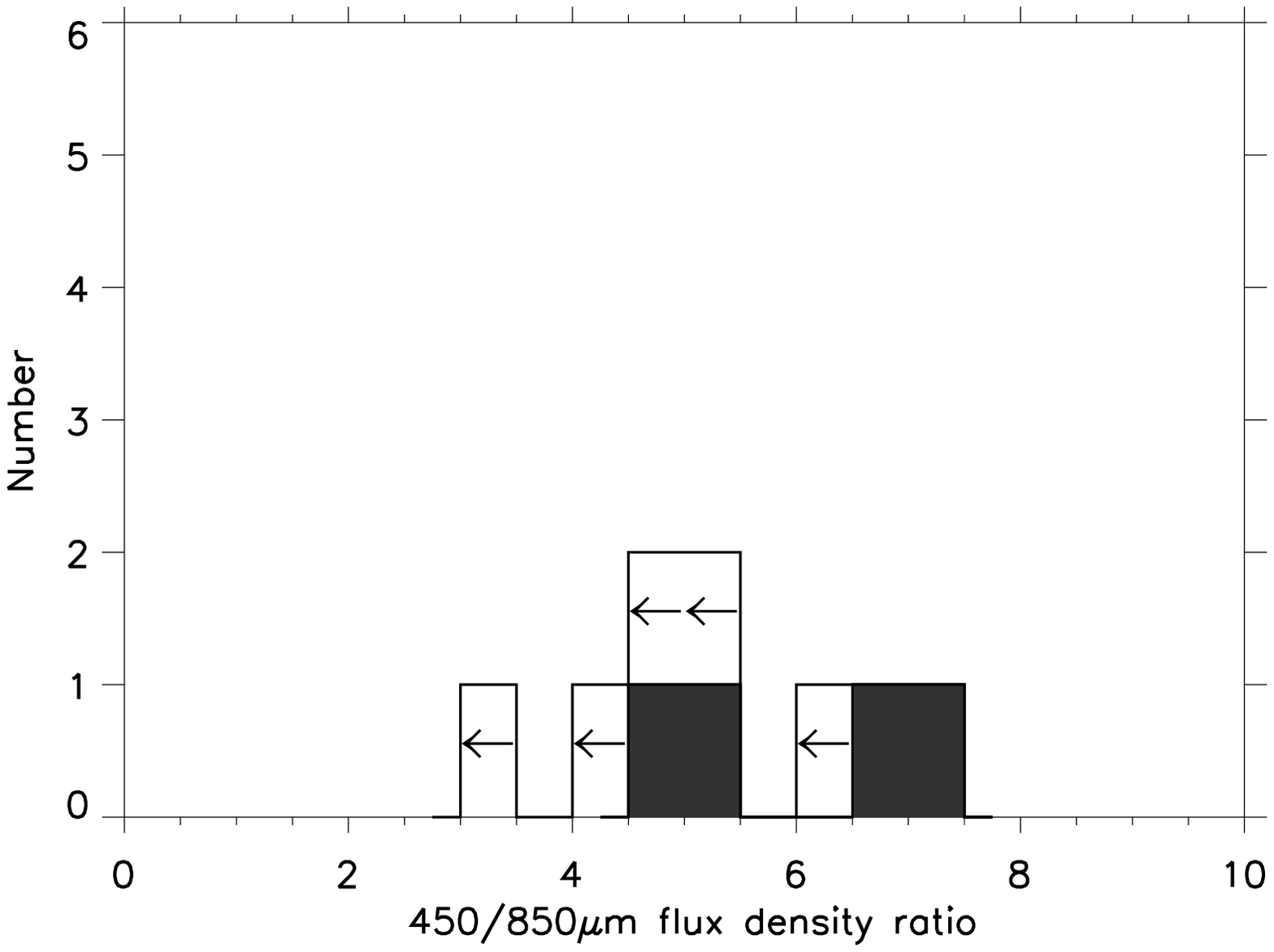,width=8.3cm,clip=}\\
\psfig{file=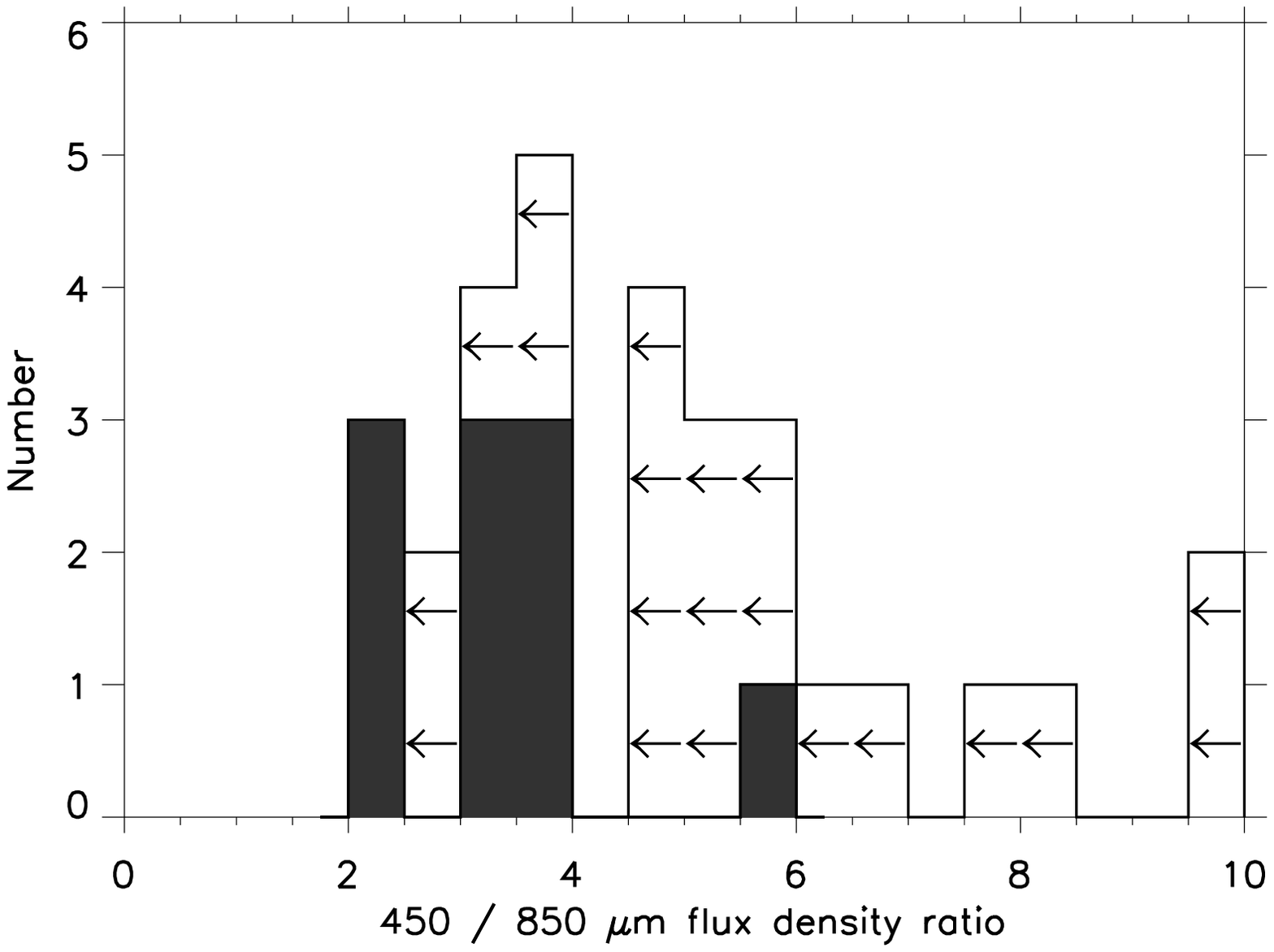,width=8.3cm,clip=}
\end{tabular}
\caption{\label{ratios} {\it Top:} The 450 to 850$\mu$m flux density ratio
as a function of redshift derived by Lutz et~al. (2001) by redshifting the
spectral energy distributions of 14 local ultra--luminous infrared
galaxies (ULIRGs). The solid line displays the median relation and the
shaded region shows the total scatter of the points. {\it Middle:}
Histograms of the 450 to 850$\mu$m flux density ratios for those sources
detected with a significance above 4$\sigma$ in the fields of the high
redshift clusters. The filled boxes represent sources detected at both
wavelengths and the open boxes represent the 4$\sigma$ upper limits for
sources undetected at 450$\mu$m. Only the measured ratios or limits are
shown; statistical and calibration errors (which can be $\gta 25$\%) are
not indicated. {\it Bottom:} A similar plot combined for sources in the
blank fields of the 8\,mJy survey (Scott et~al 2002; Fox et~al 2002) and
the SCUBA lensing survey (Smail et~al 1997, 2002). The sources detected
towards the high redshift clusters have systematically higher 450 to
850$\mu$m flux density ratios, consistent with being at lower redshifts
($z \sim 1$) than those found in blank field surveys (typically $z \gta
2$).}
\end{figure}

Lutz \etal\ \shortcite{lut01} investigated the how the 450 to 850$\mu$m
flux density ratio would vary as a function of redshift from the spectral
energy distributions of 14 local ultra--luminous infrared galaxies
(ULIRGs) with existing far--infrared and sub--mm photometry (from Klaas
\etal\ 2001).\nocite{kla01} The relation they derived is shown in
Figure~\ref{ratios}a. 450 to 850$\mu$m flux density ratios were calculated
for all of the securely detected 850$\mu$m sources in the current sample
within the sky area of the 450$\mu$m maps. These ratios are tabulated in
Table~\ref{sources} and are displayed as a histogram in
Figure~\ref{ratios}b. For 4 sources a flux density ratio is determined,
for 3 of which the ratio is high, $\gta 5$, consistent with being at
redshifts $z \lta 1$. A further 4 sources have determined upper limits on
their flux ratios, which in 3 cases suggest relatively high redshifts when
compared directly to the predictions of Figure~\ref{ratios}a, although the
presence of any cooler ($T \sim 10-15$\,K) dust components within the
galaxies would give rise to lower 450 to 850$\mu$m flux density ratios at
lower redshifts.

In Figure~\ref{ratios}c a similar histogram of flux density ratios has
been derived for sources detected in blank field surveys, using the 8\,mJy
survey \cite{sco02,fox02} and the SCUBA lensing survey (e.g. Smail et~al
1997, 2002); these provide a good spread of sources both more and less
luminous than those of the high redshift cluster sample, and so provide a
good comparison sample. This figure provides a strong visual indication
that the sources in the current sample have systematically higher 450 to
850$\mu$m flux density ratios than those in the blank field surveys, for
example considering the fraction of sources with 450 to 850$\mu$m flux
density ratios above and below 4.0 in each sample. Survival analysis
(e.g. Feigelson \& Nelson 1985) can be used to statistically investigate
datasets such as these with large numbers of upper limits. The
Kaplan-Meier estimator gives a mean 450 to 850$\mu$m flux density ratio
for the blank field selected sources of $3.0 \pm 0.2$ and even if all of
the 5 upper limits for the cluster sample are between 2.0 and
2.5\footnote{The Kaplan-Meier estimator cannot be applied to the cluster
sample since the lowest data point is an upper limit rather than a
measured value.} then the mean for that sample will be $\gta 4$. Using the
{\sc ASURV} survival analysis package \cite{lav92} the likelihood of the
sources from the cluster and blank field surveys being drawn from the same
parent population, as determined from the 450 to 850$\mu$m flux density
ratio distributions, was calculated according to 5 different survival
analysis methods based around a generalised Wilcoxin test: taking the mean
of these results, the probability of the parent samples being the same is
$\lta 1.8$\%.

To summarise, the 450 to 850$\mu$m flux density ratios of the current
sample are systematically higher than those of blank field surveys. This
is most likely due to an extra population of sources associated with the
high redshift clusters, which have higher 450 to 850$\mu$m flux because
their redshifts are significantly lower ($z \sim 1$) than sources
typically detected in blank field surveys ($z \gta 2$; e.g. Smail \etal\
2000, Fox \etal\ 2002).\nocite{sma00,fox02}

\nocite{sma02,sma97,sco02,fox02}

\subsection{Dust masses and star--formation rates}
\label{dmass}

The mass of warm dust, $M_{\rm d}$, can be calculated from the
sub-millimetre flux using the equation:

\begin{displaymath}
M_{\rm d} = \frac{S(\nu_{\rm obs}) D_{\rm L}^2}
{(1+z)\kappa_{\rm d}(\nu_{\rm rest})B(\nu_{\rm rest},T_{\rm d})}
\end{displaymath}

\noindent where $S$ is the observed flux density, $\nu_{\rm obs}$ and
$\nu_{\rm rest}$ are the observed and rest--frame frequencies, $D_{\rm L}$
is the luminosity distance, $z$ is the redshift, $\kappa_{\rm d}$ is the
mass absorption coefficient, $B$ is the black--body Planck function, and
$T_{\rm d}$ is the dust grain temperature. Here, a dust temperature
$T_{\rm d} \sim 40$\,K, and a mass absorption coefficient $\kappa_{\rm d}
= 0.067 (\nu_{\rm rest} / 250 {\rm GHz})^{\beta}$\,m$^2$\,kg$^{-1}$ with
$\beta=2$ are adopted: for a discussion of the uncertainties on these
parameters see Hughes, Rawlings and Dunlop (1997) \nocite{hug97}.  With
these properties then a source with a flux density of 10\,mJy at 850$\mu$m
would have a dust mass of $\sim 5 \times 10^8 M_{\odot}$ if it were at
redshift $z \sim 1$. If it is assumed that this dust is heated primarily
by young stars, then scaling the sub--mm luminosity to the star--formation
rate by using nearby starbursts such as M82 \cite{hug97}, the star
formation rate for a 10\,mJy 850$\mu$m source at $z=1$ is $\sim 2000
M_{\odot}$yr$^{-1}$. Thus any confirmed sub-mm emitting cluster members
are undergoing extremely active starbursts.

\section{Conclusions}
\label{concs}

SCUBA observations have been presented of the central regions of four rich
cluster of galaxies at redshifts $z \sim 1$. Compared with the number
counts determined from blank--field surveys, an excess of sub-mm sources
of a factor $\sim 3-4$ at 850$\mu$m and up to an order of magnitude at
450$\mu$m has been detected. Some fraction of this excess will be due to
the effects of gravitational lensing magnification of background sources
by the clusters; an analysis of possible source count boosting by
gravitational lensing has been carried out and demonstrates that, unless
the measured velocity dispersions greatly underestimate the cluster
masses, this could boost the source counts by at most a factor of two. The
residual excess counts are most likely directly associated with the high
redshift clusters. This conclusion is strengthened by the typically higher
450 to 850$\mu$m flux density ratios of these sources as compared with
blank--field selected sources. These higher ratios are consistent with a
significant proportion of the sources being at lower redshifts than those
of the blank field surveys (typically $z \gta 2$).

Confirmation that these sub-mm sources are indeed associated with cluster
galaxies would indicate a dramatic increase with redshift in the
proportion of strongly starbursting galaxies within clusters. This mirrors
the increasing fraction of blue, star--forming galaxies indicated by the
Butcher--Oemler effect \cite{but78}, and the recent discovery that more
distant clusters often contain an enhanced population of weak radio
sources \cite{dwa99,mor99,bes02a}. All of these results demonstrate that
high redshift clusters are still in a very active state of formation. With
a larger dataset it would be particularly interesting to investigate the
radial distribution of the sub-mm sources within the clusters, and to
investigate whether mergers between clusters, which are common at these
high redshifts, increase the proportion of starbursting galaxies.

\section*{Acknowledgements} 

I would like to thank the Royal Society for generous financial support
through its University Research Fellowship scheme. The JCMT is operated by
the Joint Astronomy Centre on behalf of the United Kingdom Particle
Physics and Astronomy Research Council (PPARC), the Netherlands
Organisation for Scientific Research and the National Research Council of
Canada. I thank Susie Scott for useful discussions about matters of SCUBA
data reduction and Rob Ivison et~al for supplying the 450$\mu$m flux
densities of the SCUBA lensing survey sample prior to publication. I
particularly thank Ian Smail, Rob Ivison and Jim Dunlop for their careful
reading of the manuscript and helpful comments. I am grateful to the
referee for a number of suggestions which have improved the manuscript.

\label{lastpage}
\bibliography{pnb} 
\bibliographystyle{mn} 

\end{document}